\documentclass[aps,showpacs,amsmath,amssymb,twocolumn,prx,longbibliography,superscriptaddress,notitlepage,floatfix,10pt]{revtex4-2}

\usepackage[dvips]{graphicx}
\usepackage{bm, bbm,ascmac,amsmath,amssymb,amsthm,mathrsfs,amsfonts,dsfont}
\usepackage{epsfig}
\usepackage{braket}
\usepackage{enumerate}
\usepackage{xcolor}
\usepackage{float}
\usepackage{algorithm}
\usepackage{algorithmic}
\usepackage{comment}
\usepackage{appendix}
\usepackage{here}
\usepackage{tabularx}
\usepackage{dcolumn}
\usepackage[caption=false]{subfig}
\usepackage{adjustbox}
\usepackage[hidelinks]{hyperref}
\usepackage{bookmark}
\usepackage{physics}
\usepackage{enumitem}

\usepackage{scalerel}

\iftrue
  \usepackage{marginnote}
\fi

\theoremstyle{plain}
\newtheorem{theorem}{Theorem}
\newtheorem{lemma}{Lemma} 

\theoremstyle{definition}
\newtheorem{definition}{Definition} 

\usepackage{todonotes}
\todostyle{holl}{color=green!30, size=\footnotesize}
\todostyle{holli}{color=green!30, inline, size=\footnotesize}
\todostyle{by}{color=blue!30, size=\footnotesize}
\todostyle{byi}{color=blue!30, inline, size=\footnotesize}

\usepackage [n, advantage, operators, sets, adversary, landau, probability,  notions, ff, mm, primitives, events, complexity, asymptotics, keys ] {cryptocode}

\begin{document}

\newfloat{protocol}{htbp}{idf} \floatname{protocol}{Protocol~}

\newfloat{resource}{htbp}{idf} \floatname{resource}{Resource~}

\newfloat{simulator}{htbp}{idf} \floatname{simulator}{Simulator~}

\newfloat{problem}{htbp}{idf} \floatname{problem}{Problem~}

\title{Verifiable blind observable estimation}

\author{Bo Yang} \email{Bo.Yang@lip6.fr} \affiliation{LIP6, Sorbonne Université, CNRS, 4 place Jussieu, 75005 Paris, France}

\author{Elham Kashefi} \email{Elham.Kashefi@lip6.fr} \affiliation{LIP6, Sorbonne Université, CNRS, 4 place Jussieu, 75005 Paris, France} \affiliation{School of Informatics, University of Edinburgh, 10 Crichton Street, EH8 9AB Edinburgh, United Kingdom}

\author{Harold Ollivier} \email{harold.ollivier@ens.fr} \affiliation{QAT, DIENS, École Normale Supérieure, PSL University, CNRS, INRIA, 45 rue d'Ulm, Paris 75005, France}

\begin{abstract}
Cryptographic verification is essential for establishing trust in quantum-computing-as-a-service.
However, a fundamental gap exists in the current verification landscape: existing efficient protocols are largely restricted to decision problems where correctness is boosted by classical majority voting.
This excludes observable estimation, the statistical task underpinning nearly all near-term quantum advantage applications.
For such tasks, current verification techniques face a prohibitive trade-off: either weak security guarantees or massive space overhead that exceeds the capacity of near-term hardware.
To resolve this, we introduce the Secure Delegated Observable Estimation (SDOE) ideal resource, the first formal cryptographic framework for trustworthy expectation-value estimation within Abstract Cryptography.
We then present the Verifiable Blind Observable Estimation (VBOE) protocol, which efficiently constructs this resource.
VBOE circumvents the limitations inherent in prior methodologies by enabling the sequential collection of samples with negligible security error, requiring zero extra qubit overhead.
By directly averaging computation rounds in classical post-processing, our protocol provides the only known path to rigorous, composable verification for the most common class of near-term quantum-classical hybrid algorithms.
This work bridges foundational cryptographic theory with practical quantum tasks, enabling the certification of quantum utility on current and near-future devices.
\end{abstract}

\maketitle

\section{Introduction\label{sec:Introduction}}

Quantum computing promises to solve problems intractable for classical machines, a capability known as quantum advantage~\cite{Shor1994Algorithms, Grover1996A, Preskill2018quantumcomputingin, Preskill2012Quantum, Lanes2025A, Huang2025The}.
Recent advances in quantum hardware~\cite{ni2023beating, gupta2024encoding,Hughes2025Trapped-ion,Ransford2025Helios} have brought this promise closer to reality, with near-term devices demonstrating potential for advantage in tasks that do not require full fault tolerance~\cite{Arute2019Quantum, Zhong2020Quantum, Kim2023Evidence, Robledo-Moreno2025Chemistry}.
However, because these devices are error-prone and often remotely accessed, they raise critical questions: How can a client ensure the privacy of a delegated quantum computation while also verifying its result, especially when the output is classically intractable to compute? 

This challenge is particularly acute for algorithms based on expectation value estimation, which underpin many near-term quantum applications such as quantum simulation~\cite{McArdle2020Quantum, cerezo2021variational} and machine learning~\cite{cerezo2021variational, Larocca2025Barren, Mhiri2025A, Cerezo2025Does}.
For these applications, establishing a rigorous method to verify the returned estimates is a prerequisite for achieving reliable quantum utility.
Yet, even in the case of honest but noisy devices, a fundamental challenge remains: is it possible for a purely classical client to efficiently and robustly verify the outcome of a task delegated to a remote server?
While particular schemes have been developed for specific problems~\cite{AaronsonPeaked2024}, no general practical solution currently exists for the broad class of near-term algorithms mentioned above.

\begin{figure*}[htbp]
    \centering
    \subfloat[Round-wise verification]{
        \includegraphics[width=0.48\linewidth]{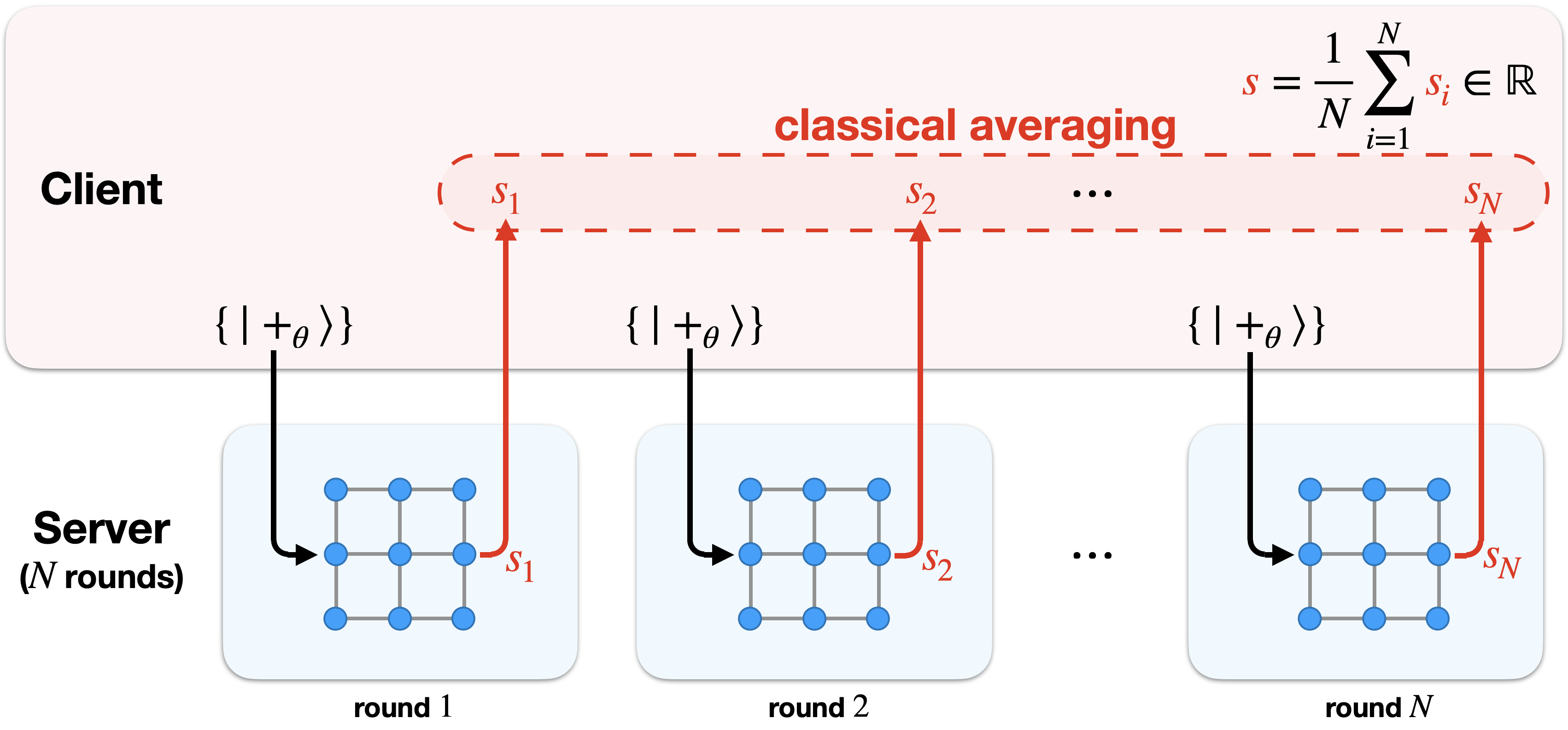}
    }
    \hfill
    \subfloat[End-to-end verification]{
        \includegraphics[width=0.48\linewidth]{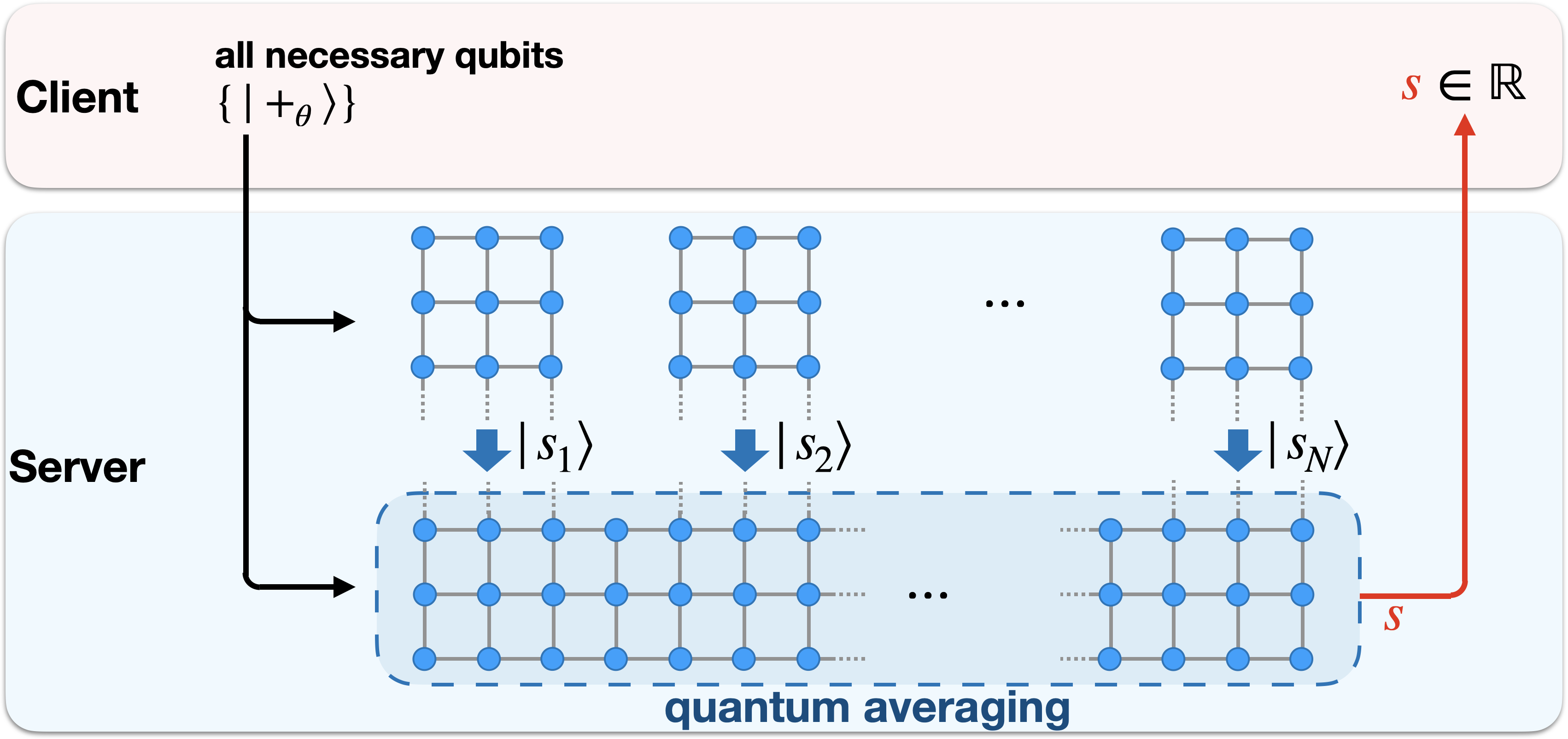}
    }
    \caption{
        (a) Schematic illustration of round-wise verification, in which the Server sends all samples directly to the Client, who then computes the empirical mean classically.
        In this setting, the security of each round is guaranteed to be inverse-polynomial in the number of rounds under conventional security analysis, implying that the entire protocol has at best inverse-polynomial security.
        (b) Schematic illustration of end-to-end verification, in which the Server returns a final empirical mean obtained via quantum averaging.
        In this case, exponential security can be achieved within the conventional security analysis; however, quantum averaging incurs substantial space overhead, which would be prohibitive for near-term quantum hardware.
        Note that in both figures, the black arrows represent quantum communication to send single qubits, and the red arrows represent the classical communication to return the final outcome from the server.
    }
    \label{fig:vbqc}
\end{figure*}

Recently, a landmark experiment~\cite{MK2025Verifiable} proposed a heuristic method where one quantum computer simulates the results of another.
While this represents a major step forward, the approach is not mathematically rigorous.
If two devices happen to share the same noise profile, they could produce identical results in a regime beyond classical simulability while being equally incorrect.
Conversely, if the devices provide different results due to differing noise effects, there is no rigorous way to determine which result is correct.

To resolve this, researchers over the last two decades have established rigorous, scalable verification techniques based on cryptographic obfuscation.
These protocols, generally referred to as Verifiable Blind Quantum Computing (VBQC)~\cite{gheorghiu2019a} are formulated as an interactive proof system.
A verifier with minimal quantum capabilities, specifically single-qubit preparation, can rigorously prove the correctness of a computation performed by a potentially untrusted and noisy server.
By randomly inserting traps while keeping the entire computation process obfuscated, the absence of deviations in the trap computations guarantees the correctness of the computation~\cite{fitzsimons2017unconditionally}.
Such techniques require the verifier to be quantumly linked to the server, a capability demonstrated in both off-chip and on-chip settings~\cite{Gustiani2025On-chip,Polacchi2025Experimental,drmota2024verifiable}.
This confirms the practicality of these schemes and their alignment with the modular scalable roadmaps of the major hardware platforms.

Nevertheless, a critical subtlety remains.
To obtain exponential confidence in the correctness of the result, one must amplify the security bound.
If the goal is the correctness of the entire quantum output, the only known solution is full fault-tolerant computing~\cite{fitzsimons2017unconditionally}.
Fortunately for decision problems (BQP), that is the most common usage of quantum computers, one can instead blindly interleave computation rounds and test rounds, containing only traps, and apply a simple post-processing, majority voting, to achieve exponential security~\cite{leichtle2021verifying} (see Protocol~\ref{proto:rvbqc} in Appendix~\ref{sec:appendix_preliminaries}).
Such a solution is then efficient as the security error is exponentially suppressed as the number of computation rounds and test rounds grows, and is unaffected by the size of the computation.

In the case of the observable estimation algorithms, the desired post-processing is instead the classical computation of the empirical average.
However, this simple change breaks the existing verification framework.
If the server sends all samples directly to the verifier to classically compute the average as depicted in Fig.~\ref{fig:vbqc}(a), there is no guarantee that individual samples were correctly computed; unlike decision problems, sampling problems lack simple classical error-detection schemes.
This results in a security guarantee that is at best inverse polynomial in the number of rounds.
To recover exponential security guarantees, the server should perform the empirical mean quantumly by processing all samples in parallel as depicted in Fig.~\ref{fig:vbqc}(b).
Then, the space complexity of the protocol increases significantly, making the approach impractical for current hardware.
This represents a fundamental gap that prevents existing verification schemes from being applicable directly to the most important class of problems targeting near-term quantum utility.

To address these limitations, we first introduce the Secure Delegated Observable Estimation (SDOE) resource.
This is a new ideal resource within the abstract cryptography (AC) framework~\cite{Maurer2011abstract} that formalises a trustworthy estimation process.
SDOE is specifically designed to handle the statistical nature of near-term tasks, ensuring the client obtains an estimate of an observable’s expectation value within a specified bias, $\epsilon$, or detects a deviation and aborts, all while preserving the confidentiality of the observable and the result.

We then present the Verifiable Blind Observable Estimation (VBOE) protocol, which efficiently constructs the SDOE resource.
VBOE resolves the aforementioned gap by enabling a security error that is negligible in the number of rounds without requiring any extra qubit overhead.
Unlike the problematic trade-offs of previous approaches, VBOE allows for the sequential collection of samples, avoiding the space blowup of parallel computation, while using a novel verification analysis to ensure the integrity of the empirical average.
Our main result is a rigorous proof that VBOE constructs SDOE with composable security, providing the first efficient framework for verifying observable estimation tasks on untrusted devices.

As a consequence, our protocol directly addresses the growing demand for verifiable quantum advantage in the NISQ era.
Its absence of overhead provides a path to polynomial-time verification with exponentially small soundness error for tasks performed on current state-of-the-art quantum hardware~\cite{Babbush2025The}.
This bridges the gap between foundational cryptographic theory and the heuristic methods recently used in milestone experimental demonstrations of quantum advantage~\cite{Abanin2025Observation, MK2025Verifiable}.

\section{Verifying observable estimation}
\label{sec:Verifying_observable_estimation}

We define the Secure Delegated Observable Estimation (SDOE) Resource~\ref{res:sdoe} and show that it can be constructed by the proposed Verifiable Blind Observable Estimation (VBOE) Protocol~\ref{proto:vboe} to negligible error within the AC framework.

\subsection{Observable estimation problems}
\label{sec:Definition_of_observable_estimation}

A generic observable estimation problem consists of computing $\operatorname{Tr}\left[\rho O\right]$ for some state $\rho$ and observable $O$ up to an additive error $\epsilon > 0$. Without loss of generality, we can always assume that $O$ is a coarse-grained measurement of $n$-qubits in the $\ket\pm$ basis for $n$ sufficiently large. This is because it suffices to absorb the basis change for the eigenspace of $O$ into $\rho$. As a result, an observable estimation problem is completely specified by $\mathsf{C}\in\mathfrak{C}$, a computation that produces $\rho$ from some fiducial state, say $\ket +^{\otimes m}$ for some $m \geq n$. A further simplification that we will adopt in the remainder of this paper is to assume that $O$ is indeed a binary observable, e.g. $X=|+\rangle\langle+| - |-\rangle\langle-|$ with eigenvalues $-1$ and $1$, so that the observable estimation consists of producing a single qubit state $\rho$ and measuring it in the $\ket \pm$ basis. We will argue in Section~\ref{sec:discussion} that our protocol and result can be straightforwardly extended to bounded non-binary observables.

A naive estimation procedure works by sampling outcomes $y_{i}=2Y_{i}-1\in\{-1,1\}$ where $Y_{i} \sample \mathcal B(p)$ and $\displaystyle p = \operatorname{Tr}\left[\rho |+\rangle\langle+|\right]=\frac{1+\operatorname{Tr}\left[\rho O\right]}{2}$ is the probability of obtaining 1 in the measurement of $\rho$ according to observable $X$, while $\mathcal{B}(p)$ is the Bernoulli distribution; and then by computing the empirical average of $N_{c}$ samples.
\begin{equation} \mu = \frac{1}{N_{c}} \sum_{i=1}^{N_{c}} {y}_{i}.
\end{equation} Using Hoeffding's bound (Lemma~\ref{lemma:ineq_hoeffding}), one can assess the performance of such a procedure:
\begin{equation}
\begin{split} \operatorname{Pr}\left[|\mu - \operatorname{Tr}\left[\rho O\right]|\geq\epsilon\right] \leq 2\exp\left(-\frac{\epsilon^{2}}{2}N_{c}\right).
\end{split}
\end{equation} It states that for fixed $\epsilon$, the probability of the estimator being further away than $\epsilon$ from the true value $\operatorname{Tr}\left[\rho O\right]$ is negligible in $N_{c}$, the number of collected samples. This motivates the following definition:
\begin{definition}[$\left(\epsilon, \delta\right)$-Observable Estimation] Given an observable $O$, a reference state $\rho$, a protocol $(\epsilon, \delta)$-estimates $\operatorname{Tr}\left[\rho O\right]$ if the protocol outputs an estimate $o$ that satisfies
 \begin{equation}\label{eq:epsilon_inc} \operatorname{Pr}\left[|o - \operatorname{Tr}\left[\rho O\right]| \geq \epsilon\right]\leq \delta.
 \end{equation} Above, $\epsilon>0$ is the allowed bias and $\delta>0$ is an upper bound on the failure probability for obtaining an estimate within the allowed bias.
\end{definition}

\subsection{Secure delegated observable estimation (SDOE)}
\label{sec:Secure_delegated_observable_estimation}

To formalise security for observable estimation problems, we define an ideal resource that has perfect blindness and always returns an estimate of $\operatorname{Tr}\left[\rho O\right]$ within bias $\epsilon$ in the form of Resource~\ref{res:sdoe}.

\begin{resource}[htbp]
    \caption{\raggedright Secure Delegated Observable Estimation (SDOE)}
    \label{res:sdoe}
    \begin{algorithmic}[0] \STATE \textbf{Public information:} $\mathfrak{C}$ a computation class; $N_{c}, N_{t} \in \mathbb{N}$; a security parameter $\omega>0$; and the allowed bias $\epsilon>0$.
        
        \STATE \textbf{Client's interface:} The target computation $\mathsf{C}\in \mathfrak{C}$ to produce the single qubit state $\rho$ to be measured by $O = X$.
        
        \STATE \textbf{Server's interface:}
        \begin{enumerate}
            \item The interface is filtered so that when $e=0$, the interface does not send any information nor take inputs.
            \item For $e=1$, the Resource receives a quantum state $\sigma$ and $F$, a list of instructions so that the resource produces $s \in \mathbb R \cup \{\mathsf{Abort}\}$.
        \end{enumerate}
        
        \STATE \textbf{Processing by the Resource:}
        \begin{enumerate}
            \item If $e=0$, it sets $\displaystyle o = \frac{1}{N_{c}} \sum_{i = 1}^{N_{c}}y_{i}$ with $y_{i}=2Y_{i}-1$ where $Y_{i} \sample \mathcal{B}(p)$, sampled from a Bernoulli distribution $\mathcal{B}(p)$ with $p = \operatorname{Tr}\left[\rho |+\rangle\langle+|\right]$.
            \item If $e=1$, it computes $s$ using the transmitted state $\sigma$ and $F$.
            \item If $s = \mathsf{Abort}$ it forwards $o=\mathsf{Abort}$ to the Client.
            \item If $|s - \operatorname{Tr}\left[\rho O\right]| \geq \epsilon$ it sets $o = \mathsf{Abort}$ and forwards it to the Client.
            \item Otherwise it directly forwards $s$ to the Client.
        \end{enumerate}
    \end{algorithmic}
\end{resource}

Here, $e\in\{0,1\}$ is a flag controlling whether the Server's interface filter is activated or not. Whenever $e=0$, the ideal resource samples the estimator of $\operatorname{Tr}\left[\rho O\right]$ and returns its value if it is within $\epsilon$ of the true expectation value. When the Server asks for full access ($e=1$), the Server receives at most the permitted leakage, i.e. essentially $\mathfrak{C}$ corresponding to all the observable estimation problems that the resource can handle. It also recieves the parameters $N_{c}$, $N_{t}, w$ and $\epsilon$. The Server is then allowed to send a deviation to be applied by the Resource. It takes the form of a quantum state $\sigma$ and a classical list of quantum and classical instructions that produce either a real number or $\mathsf{Abort}$. If the produced scalar is within $\epsilon$ of $\operatorname{Tr}\left[\rho O\right]$, then the resource sends the scalar to the Client. Otherwise, it sends $\mathsf{Abort}$.

\begin{figure}[htbp] \centering \includegraphics[width=\linewidth]{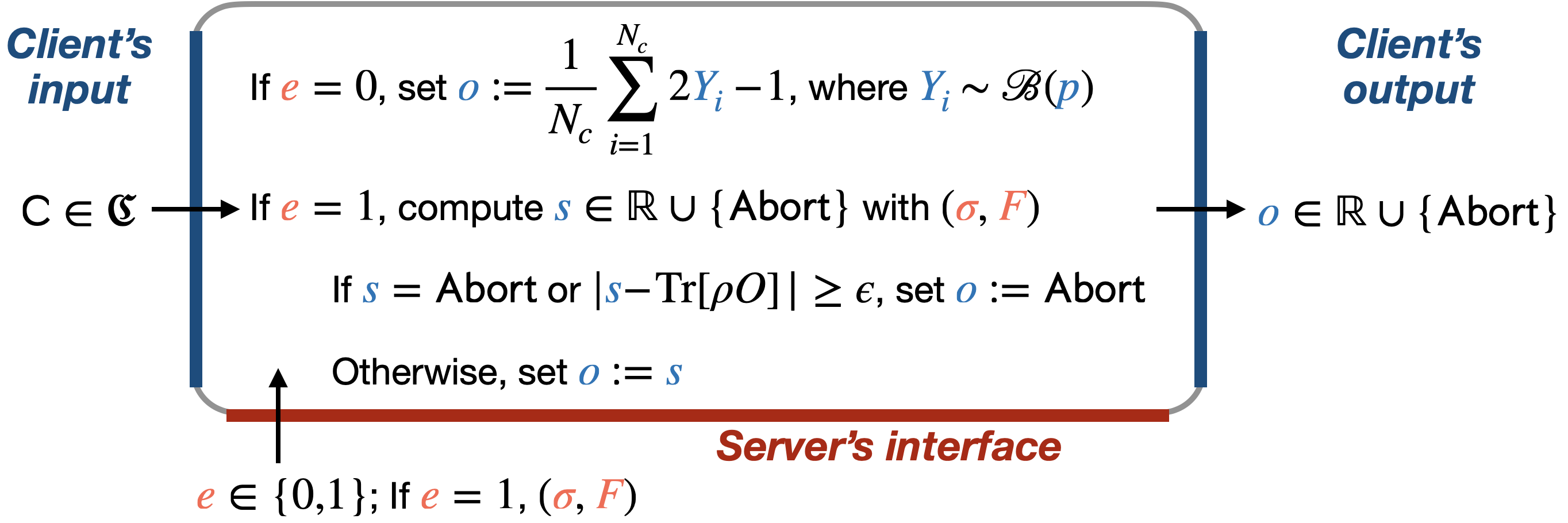}
  \caption{ Schematic illustration of the SDOE resource (Resource~\ref{res:sdoe}). The bottom edge of the outer rectangle serves as an interface to the Server. The left edge of the outer rectangle takes inputs from the Client, and the right edge returns outputs to the Client. The variables and equations coloured in blue represent the values generated in the SDOE resource, while those coloured in light red represent the values received at the Server's interface. }
  \label{fig:sdoe}
\end{figure}

This definition corresponds to the intuitive notion of secure delegated observable estimation: a malicious Server can only learn the class of observable estimation problems that the resource can tackle, and is only able to influence the result so long as it remains within $\epsilon$ of the true expectation value. It also inevitably has the ability to force the protocol to $\mathsf{Abort}$. The pictorial representation of the SDOE resource is provided in Fig.~\ref{fig:sdoe}.

\subsection{Verifiable blind observable estimation (VBOE)\label{sec:Verification_protocol_observable_estimation}}

Considering the SDOE resource defined above, we propose the VBOE protocol to concretely implement this functionality. It is based on the same sequential execution of test and computation rounds as in the RVBQC (See appendix, Protocol~\ref{proto:rvbqc})~\cite{leichtle2021verifying}, with post-processing and acceptance criteria tailored to observable estimation problems. Here we assume that $\mathfrak{C}$ is a computation class that corresponds to all measurement patterns that can be executed on a graph $G = (V,E)$ with a given flow $f$.

\begin{protocol}[htbp]
    \caption{\raggedright Verifiable Blind Observable Estimation (VBOE)}
    \label{proto:vboe}
    \begin{algorithmic}[0] \STATE \textbf{Inputs from Client:} The target computation $\mathsf{C}\in \mathfrak{C}$ that produces $\rho$ and allows to measure $\operatorname{Tr}\left[\rho O\right]$ with $O = X$.
        
        \STATE \textbf{Protocol:}
        \begin{enumerate}
        \item The Client randomly samples indices in $[N]$ for $N = N_{c} + N_{t}$ to indicate the locations of test and computation rounds. Let $\mathrm{S}_{\mathsf{T}}$ (resp. $\mathrm{S}_{\mathsf{C}}$) be the index set of test (resp. computation) rounds.
        \item For $i \in \mathrm{S}_{\mathsf{T}}$, the Client constructs a test round following the same procedure as for the test rounds of the RVBQC (Step 2 of Protocol~\ref{proto:rvbqc}).
        \item Each round, computation or test, is then delegated to the Server using UBQC (Protocol~\ref{proto:ubqc}).
        \item Upon receiving and decoding the result of the computation round $i \in \mathrm{S}_{\mathsf{C}}$, the Client assigns the result to $\tilde y_{i}\in\{-1,1\}$.
        \item Upon receiving the measurement results of test rounds, the Client checks that all the traps have output the expected outcome. If it is the case, the test passed. Otherwise, it failed.
        \item If less than $\omega N_{t}$ test rounds failed, it sets $\displaystyle \tilde o = \frac{1}{N_{c}} \sum_{i \in \mathrm{S}_{\mathsf{C}}}\tilde y_{i}$ as the result, otherwise it sets it to $\tilde o = \mathsf{Abort}$.
        \end{enumerate}
    \end{algorithmic}
\end{protocol}

\begin{figure}[htbp] \centering \includegraphics[width=\linewidth]{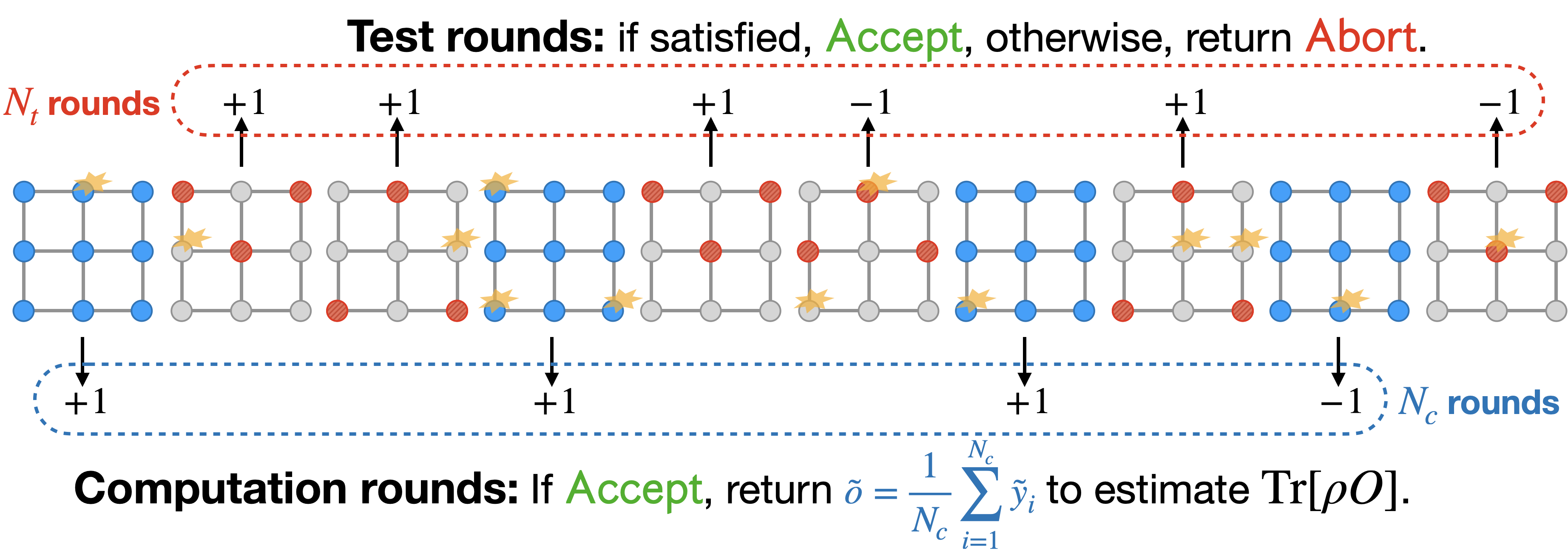}
    \caption{The schematic illustration of the VBOE protocol. The test rounds have only trap qubits and dummy qubits, shown in red and grey circles, respectively. The computation rounds have only the computation qubits shown in blue circles.}
    \label{fig:vboe}
\end{figure}

The main modification in VBOE from RVBQC is the post-processing of computation rounds. RVBQC applies classical majority vote over multiple repeated runs to amplify the probability of choosing a correct outcome, while VBOE computes the empirical average over $N_{c}$ outcomes $\{\tilde{y}_{i}\}_{i \in \mathrm{S}_{\mathsf{C}}}$. To ensure that the returned value stays within the allowed bound $\epsilon$, the threshold $\omega$ needs to be set at the appropriate value. More precisely, we will see in the security proof that it is set in a way that the concentration of probabilities ensures honest executions are always accepted, while deviations that could generate a result too far away from the true value are rejected. The schematic illustration of the VBOE protocol is depicted in Fig.~\ref{fig:vboe}.

\subsection{Concrete construction of the SDOE resource}
\label{sec:Proof_of_composability}

The VBOE protocol combines blindness and verifiability and constructs the SDOE resource within negligible error in the AC framework:
\begin{theorem}[Composable Security of VBOE] \label{thm:vboe} Let $\mathfrak{C}$ be a class of observable estimation problems that can be estimated using an MBQC pattern on a fixed graph $G$ with a given flow $f$ and chromatic number $K$. Let $N_{c}, N_{t},\in \mathbb{N}$, let $\epsilon, \omega$ be constants such that $0 \leq K\omega < \epsilon$. Then, the VBOE protocol (Protocol~\ref{proto:vboe}), with $N_{c}$ computation rounds and $N_{t}$ test rounds, $\delta$-constructs the SDOE resource. For a constant ratio $N_{c}/N_{t}$, $\delta$ is negligible in $N_{c}$.
\end{theorem}

Following abstract cryptography, to prove this theorem, we need to upper bound the distinguishing advantage between the VBOE protocol and the SDOE resource in the honest (Correctness proof) and malicious (Security proof) settings (see Definition~\ref{def:Construction_of_Resources} in Appendix~\ref{sec:appendix_preliminaries}).

\begin{proof}[Correctness] The proof of correctness relies on the composability of the UBQC protocol. As apparent in Protocol~\ref{proto:vboe}, each round is delegated to the server using UBQC (Protocol~\ref{proto:ubqc} in Appendix~\ref{sec:appendix_preliminaries}). Because UBQC perfectly constructs the Blind Delegated Quantum Computation (BDQC) resource (Resource~\ref{res:bdqc} in Appendix~\ref{sec:appendix_preliminaries}), we can instead perform the proof of correctness using a hybrid protocol where each instantiation of UBQC is replaced by a call to BDQC.

As a result, each outcome $y_{i}$ for the computation rounds $i\in \mathrm{S}_{\mathsf{C}}$ is processed through $y_{i}=2Y_{i}-1$, where $Y_{i}$ is sampled from the Bernoulli distribution $\mathcal{B}(p)$ with probability $\displaystyle p = \frac{1 + \operatorname{Tr}\left[\rho O\right]}{2}$. This ensures that the produced empirical average $\displaystyle \tilde\mu =\frac{1}{N_{c}}\sum_{i\in \mathrm{S}_{\mathsf{C}}} \tilde y_{i}$ is obtained from the same probability distribution as the one used to define the ideal resource. Hence, whenever the ideal resource and the protocol both output the estimated value or both output $\mathsf{Abort}$, their outputs coincide. Consequently, the only distinguishing advantage stems from the two setups having different $\mathsf{Abort}$ probabilities.

Indeed, because the test rounds in the protocol are executed perfectly, they consistently give the correct outcome, and the protocol never aborts. For the ideal resource, this is not the case. Whenever the estimator is further away than $\epsilon$ from $\operatorname{Tr}\left[\rho O\right]$, the ideal resource returns $\mathsf{Abort}$. The probability of such an event happening is upper bounded, using Hoeffding's bound, by $\displaystyle 2\exp\left(-\frac{\epsilon^{2}}{2} N_{c}\right)$. We can thus conclude that, for $\epsilon$ fixed, the distinguishing advantage in the honest case is a negligible function of $N_{c}$.
\end{proof}

\begin{proof}[Security] The security relies heavily on the composability of UBQC. We start by constructing a simulator that we attach to the Server's interface of the ideal resource. Its purpose is to generate plausible transcripts and help the ideal resource in returning $o\in\mathbb{R}\cup\{\mathsf{Abort}\}$ to the Client's interface so that any distinguisher will be unable to tell apart this situation from running the Client's part of the concrete Protocol~\ref{proto:vboe}. This simulator is easily constructed from the simulator designed to prove the security of UBQC.

First, it sets $e=1$. It then prepares EPR pairs for each qubit that the Server is supposed to receive in Protocol~\ref{proto:vboe}. It sends all half EPR pairs to the Server, instructs random measurement angles, and retrieves alleged measurement outcomes. It then forwards the second half of the EPR pairs, the chosen angles and received bits to the ideal resource. It also samples at random indices within $[N_{c} + N_{t}]$ to define the sets $\mathrm{S}_{\mathsf{C}}$ and $\mathrm{S}_{\mathsf{T}}$. It decides which type of test round is associated with each $i\in \mathrm{S}_{\mathsf{T}}$ and passes this to the ideal resource.

Following the security proof of UBQC, the information passed per round, together with the nature of the round, either computation or test, is sufficient for the ideal resource to generate per-round measurement results following the same, possibly deviated, probability distributions as the ones obtained when running Protocol~\ref{proto:vboe}. From the outcomes of computation rounds, the ideal resource is then instructed to compute the empirical average $o$ and from the test rounds to check that no more than $\omega N_{t}$ failed, in which case it sets $o = \mathsf{Abort}$. Using the computed value $o$, the ideal resource then performs the steps 3, 4, and 5 mentioned in its definition (Resource~\ref{res:sdoe}) and aimed at ensuring that the returned $o$ is within $\epsilon$ of $\operatorname{Tr}\left[\rho O\right]$.

To proceed further with the proof, we note that the perfect blindness of UBQC ensures that the value of $s$ computed by using the quantum state $\sigma$ provided by the simulator, i.e. the half EPR pairs, and the classical instructions follows the same distribution as $\tilde o$ in Protocol~\ref{proto:vboe}. The only difference arrives later when the additional check $|s-\operatorname{Tr}\left[\rho O\right]|\geq\epsilon$ is performed by the resource and possibly rejects when the protocol would have accepted.

As a result, the distinguishing advantage for telling apart the concrete protocol from the ideal resource is bounded by the total variation distance between the probability distributions of $s$ and $o$, or equivalently between $o$ and $\tilde{o}$. Because, conditioned on $o, \tilde{o} \in \mathbb R$ or $o, \tilde{o} = \mathsf{Abort}$, the distributions of $o$ and $\tilde{o}$ are the same, the total variation distance reduces to:
\begin{equation}
\begin{split} &|\operatorname{Pr}[o = \mathsf{Abort}] - \operatorname{Pr}[\tilde o = \mathsf{Abort}]| \\ &\quad\quad = \operatorname{Pr}[\tilde o \neq \mathsf{Abort} \wedge |\tilde o -\operatorname{Tr}\left[\rho O\right]| \geq \epsilon ].
\end{split}
\end{equation}

This probability can be upper-bounded in 4 steps:
\begin{enumerate}
    \item we upper bound the probability that the computation rounds provide an empirical average that is more than $\gamma_{1}$ away from $\operatorname{Tr}\left[\rho O\right]$, with $\gamma_{1} > 0$;
    \item given that $\tilde o \neq \mathsf{Abort}$, we then upper-bound the probability that a large number of computation rounds, say $(K\omega + \gamma_{2}) N_{c}$ for $\gamma_{2} > 0$, have been attacked by the server;
    \item we recognise that attacking a fraction $\phi$ of computation rounds yields a deviated empirical average that is not away by more than $2\phi$ from the non-deviated one, due to the binary nature of the observable $O$, taking either $-1$ or $1$ for each round;
    \item we then notice that the probabilities in steps 1 and 2 above are negligible functions of $N_{c}$ and $N_{t}$. Thus, if we set $\gamma_{1} + 2(K\omega +\gamma_{2}) < \epsilon$, we have $\operatorname{Pr}\left[\tilde o \neq \mathsf{Abort} \wedge |\tilde o -\operatorname{Tr}\left[\rho O\right]| \geq \epsilon\right]$ negligible in $N_{c}$ and $N_{t}$, as it is upper bounded by the sum of the probabilities of step 1 and 2 as a result of the union bound.
\end{enumerate} This allows us to conclude that the distinguishing advantage for such a set of parameters is negligible in $N_{c}$ and $N_{t}$, thereby proving the security of the VBOE protocol, and, combined with its correctness, proving Theorem~\ref{thm:vboe}.

First, it is straightforward to see that the probability at step 1 is upper-bounded by $\displaystyle 2\exp\left(-\frac{\gamma_{1}^{2}}{2} N_{c}\right)$ using Hoeffding's inequality.

The probability of step 2 is upper-bounded in the following way.
First, because UBQC reduces the attack by the Server to a convex combination of Pauli deviations before the measurements, we can group the deviation strategies by $m$, the number of attacked rounds.
Now, because the location of test rounds and computation rounds are random, the number of affected computation rounds $Z$ follows a hypergeometric distribution with parameters $N_{c} + N_{t}$ for the total number of items, $m$ for the number of marked items and $N_{c}$ for the number of samples.
Similarly, the number of affected test rounds, denoted by $X$, follows a hypergeometric distribution with the roles $N_{c}$ and $N_{t}$ swapped.

Let $Y$ be the number of failed test rounds.
The idea is now to upper-bound the probability $\operatorname{Pr}[Z\geq (K\omega + \gamma_{2})N_{c}, Y \leq \omega N_{t}]$ using the following fact that
\begin{equation}\label{eq:max_m_two-cases}
\begin{split} 
    &\max_{m} \operatorname{Pr}\left[Z\geq (K\omega + \gamma_{2})N_{c}, Y \leq \omega N_{t}\right] \\ 
    &= \max \left\{ \max_{m \leq m_{0}}\operatorname{Pr}\left[Z\geq (K\omega + \gamma_{2})N_{c}, Y\leq \omega N_{t}\right], \right. \\ 
    &\quad\quad\quad\quad \left. \max_{m > m_{0}}\operatorname{Pr}\left[Z\geq (K\omega + \gamma_{2})N_{c}, Y\leq \omega N_{t}\right] \right\} \\ 
    &< \max_{m \leq m_{0}}\operatorname{Pr}\left[Z \geq (K\omega + \gamma_{2})N_{c}\right] + \max_{m > m_{0}}\operatorname{Pr}\left[Y\leq \omega N_{t}\right],
\end{split}
\end{equation} where $\displaystyle m_{0} = \left(K\omega+\frac{\gamma_{2}}{2}\right)(N_{c}+N_{t})$.

Within this setting, for a given value of $m \leq m_{0}$, the tail bound for the hypergeometric distribution gives
\begin{equation}\label{eq:boundZ}
\begin{split} & \operatorname{Pr}[Z \geq (K\omega + \gamma_{2})N_{c}] \\ & \leq \exp \left( -2 \left( (K\omega+\gamma_{2}) - \frac{m}{N_{c}+N_{t}} \right)^{2} N_{c}\right) \\ & \leq \exp \left( -\frac{\gamma_{2}^{2}}{2}N_{c} \right).
\end{split}
\end{equation}

To bound $\operatorname{Pr}\left[Y \leq \omega N_{t}\right]$ with $m > m_{0}$, recall that test rounds are defined as Protocol~\ref{proto:rvbqc} in RVBQC and that a given deviation on a test round is detected with probability at least $1/K$, where $K$ is the chromatic number of the underlying graph $G$. This means that conditioned on $X$, the number of failed test rounds $Y$ is lower bounded in the usual stochastic order by a binomial distribution with $X$ samples and average $1/K$. This implies that, with $m > m_{0}$,
\begin{equation}\label{eq:boundY}
\begin{split}
    &\operatorname{Pr}\left[Y\leq \omega N_{t}\right] \\
    & = \operatorname{Pr}\left[Y\leq \omega N_{t}, X \leq N_{t} \left(\frac{m_{0}}{N_{c} + N_{t}} - \frac{\gamma_{2}}{4}\right) \right] \\
    &\quad + \operatorname{Pr}\left[Y\leq \omega N_{t}, X > N_{t} \left(\frac{m_{0}}{N_{c} + N_{t}} - \frac{\gamma_{2}}{4}\right) \right] \\
    & = \operatorname{Pr}\left[Y\leq \omega N_{t}, X \leq N_{t} \left(K\omega + \frac{\gamma_{2}}{4}\right)\right] \\
    &\quad + \operatorname{Pr}\left[Y\leq \omega N_{t}, X > N_{t} \left(K\omega + \frac{\gamma_{2}}{4}\right)\right] \\
    & < \operatorname{Pr}\left[X \leq N_{t} \left(K\omega + \frac{\gamma_{2}}{4}\right)\right] \\
    &\quad + \operatorname{Pr}\left[Y\leq \omega N_{t} ~\Big|~ X = N_{t} \left(K\omega + \frac{\gamma_{2}}{4}\right)\right] \\
    & < \exp\left( -\frac{\gamma_{2}^{2}}{8} N_{t}\right) + \exp \left( - \frac{\gamma_{2}^{2}}{32K^{2}} N_{t} \right).
\end{split}
\end{equation}

Because both Equations~\eqref{eq:boundZ} and~\eqref{eq:boundY} provide bounds that are negligible in $N_{c}$ and $N_{t}$, this shows that the distinguishing advantage provided by the SDOE resource rejecting more often than the VBOE protocol is indeed negligible in $N_{c}$ for a fixed ratio $N_{c} / N_{t}$. Hence, we conclude that VBOE is constructing SDOE with negligible error in $N_{c}$ provided that $K\omega$ is below and bounded away from $\epsilon$ by a constant, so the positive constants $\gamma_{1}$ and $\gamma_{2}$ can be set such that $\gamma_{1} + 2(K\omega + \gamma_{2}) < \epsilon$.
\end{proof}

\section{Discussion}
\label{sec:discussion}

We have introduced the Verifiable Blind Observable Estimation (VBOE) protocol, which extends the capabilities of verifiable blind delegated quantum computation to the critical task of observable estimation. Our main result, Theorem~\ref{thm:vboe}, establishes that VBOE achieves composable security with exponentially negligible error and polynomial-time execution, addressing a longstanding gap in the verification of near-term quantum algorithms. 

\subsection{Enabling security in the quantum utility regime }

Our primary contribution is the extension of the available toolkit for the verification of quantum computations in the Abstract Cryptography framework. This enables embedding observable estimation in a broader, possibly hybrid, secure computational task while not having to reprove its security in this wider context. This is achieved by introducing a new conceptual tool: an ideal resource (SDOE) that formally incorporates bounded estimation error, reflecting the statistical and possibly noisy nature of near-term algorithms. This extension is not merely a technicality; it provides the only known path to achieving efficient polynomial-time verification with an exponentially small soundness error for this broad and practical class of computations. The proposed VBOE protocol is the first to construct this new resource and to outline a path to realising it in practice.

This unlocks the strongest composable security model for a vast set of near-term applications, whose security analysis was previously out of reach. For instance, in recent works, observable estimation problems, such as out-of-time-order correlations, have been put forward as milestone experiments because their quantum advantage is verifiable using another quantum hardware or natural quantum system, while lying beyond the reach of known classical simulation methods~\cite{Abanin2025Observation,MK2025Verifiable}. However, even when verification is attempted using another quantum device, the reliability of the outcome remains conceptually unresolved: without a principled cryptographic framework, the trustworthiness of the devices involved cannot be established a priori. Our framework directly addresses this gap by providing a composable, device-agnostic verification protocol for observable estimation outcomes without requiring trust in any particular quantum prover.

Our framework also provides a formal abstract cryptography security guarantee for recent applications that make use of quantum verification protocols. In particular, Inajetovic et al.~\cite{Inajetovic2025Verifiable} introduce a verifiable end-to-end delegated variational quantum algorithm (MB-DVQA) by invoking a variant of RVBQC as a subroutine to ensure the correctness of individual optimisation steps. While their work rigorously establishes the correctness and verifiability of the algorithmic outcome, a composable cryptographic security analysis of the overall construction is left open. Using our result, the MB-DVQA protocol can be analysed at the level of cryptographic resources, relying on the composable security of VBOE for delegated observable estimation tasks that were previously unavailable.

\subsection{Low implementation overhead}

A key practical advantage of VBOE is its low implementation overhead, which sets it apart from previous approaches. Conventional methods for verifying observable estimation with exponential security, such as those based on RVBQC, would typically need to verify both the acquisition of measurement outcomes and the empirical average. This would either blow up the space overhead if the outcomes are acquired in parallel, or require long-term memory to store the partial sum of outcomes if the outcomes are acquired sequentially. In both cases, we would need a fault-tolerant quantum computer.  

In contrast, the VBOE protocol is designed to preserve the structure of the original target computation generating a single measurement outcome per round. By letting the client average over the outputs of the computation rounds, VBOE suppresses the circuit overhead of previous approaches while maintaining the security guarantees. As a result, VBOE offers a significantly more practical verification strategy for observable estimation tasks, particularly in regimes where quantum resources are limited.

We envisage that applying our method on such platforms would enable end-to-end verification for more meaningful computational tasks, beyond proof-of-principle experiments. Following the first experimental demonstration of UBQC on photonic devices~\cite{Barz2012Demonstration}, recent work has reported the successful implementation of RVBQC on trapped-ion platforms~\cite{drmota2024verifiable}. With the rapid progress of quantum hardware, a growing number of high-quality quantum devices with remote access are becoming available~\cite{Hughes2025Trapped-ion, Ransford2025Helios}. Building on these developments, the proposed VBOE protocol can likewise be executed on existing quantum platforms, where more practically relevant applications based on observable estimation are within reach. This represents a significant step toward bridging theoretical protocols and practical, device-level implementations of verifiable delegated quantum computation.

\subsection{Compatibility with quantum error mitigation}

By its very nature as a protocol for verifying expectation values, VBOE is naturally compatible with near-term and early fault-tolerant error-suppression techniques. A key open question is how to securely integrate quantum error mitigation (QEM)~\cite{li2017efficient,temme2017error,endo2018practical,yang2022efficient,endo2021hybrid,cai2023quantum}, which aims to recover noise-free expectation values through classical post-processing with limited quantum overhead. Integrating QEM with VBOE has the potential to enhance the noise robustness of the protocol by reducing abort probabilities, while simultaneously enabling credible error mitigation with explicit cryptographic guarantees.

Another promising direction lies in combining multi-party variants of VBOE with device-efficient near-term techniques, such as circuit cutting and hybrid tensor-network methods~\cite{peng2020simulating,yuan2021quantum,Harada2025densitymatrix,Yang2025Resource-efficient}. Such a synergy could lead to a secure quantum-classical framework in which large-scale simulation tasks are decomposed into smaller quantum and classical subroutines and distributed across multiple remote parties.

\begin{acknowledgments} 
B.Y. acknowledges the insightful and fruitful discussions with Dominik Leichtle and Jinge Bao from the University of Edinburgh, and Eliott Mamon from Sorbonne Université. 
All authors received funding from the ANR research grants ANR-21-CE47-0014 (SecNISQ), ANR-22-PNCQ-0002 (HQI).
\end{acknowledgments}

\appendix

\section{Concentration inequalities\label{appendix:Concentration_inequalities}}

\begin{lemma}[Hoeffding's inequality]
    \label{lemma:ineq_hoeffding}
    Let $X_{1},\dots,X_{n}$ be $n\in\mathbb{N}$ independent random variables such that $a_{i} \le X_{i} \le b_{i}$ almost surely for each $i\in[n]$.
    Let $\displaystyle S_{n} = \sum_{i=1}^{n} X_{i}$.
    Then for any $t > 0$,
    \begin{equation}
    \begin{split}
        \operatorname{Pr}[S_n - \mathbb{E}[S_{n}] \le -t]
        &\leq
        \exp\left(
            -\frac{2 t^{2}}{\sum_{i=1}^{n} (b_{i} - a_{i})^{2}}
        \right), \\
        \operatorname{Pr}[S_n - \mathbb{E}[S_{n}] \ge t]
        &\leq
        \exp\left(
            -\frac{2 t^{2}}{\sum_{i=1}^{n} (b_{i} - a_{i})^{2}}
        \right).
    \end{split}
    \end{equation}
\end{lemma}

\begin{definition}[Hypergeometric distribution]
    Let $N, K, n \in \mathbb{N}$ with $0 \leq n, K \leq N$. 
    A stochastic variable $X$ is said to follow the hypergeometric distribution, denoted as $X \sim \mathrm{Hypergeometric}\left(N,K,n\right)$, if its probability mass function is described by
    \begin{equation}
    \begin{split}
        \mathrm{Hypergeometric}\left(N,K,n\right)(k)
        & = \frac {\displaystyle \binom{K}{k}\binom{N-K}{n-k}}
                {\displaystyle \binom{N}{n}}.
    \end{split}
    \end{equation}
    One possible interpretation is to see $X$ as the number of marked items when choosing $n$ items from a set of size $N$ containing $K$ marked items, without replacement.
\end{definition}

\begin{lemma}[Concentration  for the hypergeometric distribution]
    \label{lemma:ineq_tail_hypergeometric}
    Let $X \sim \mathrm{Hypergeometric}\left(N,K,n\right)$ be a random variable and $\displaystyle 0 < t < \frac{K}{N}$.
    It then holds that
    \begin{equation}
    \begin{split}
        \operatorname{Pr}\left[X \leq\left(\frac{K}{N}-t\right) n\right] \leq \exp \left(-2 t^{2} n\right).
    \end{split}
    \end{equation}
    As a corollary, we obtain the tail inequality
    \begin{equation} \label{eq:ineq_tail_hypergeometric_leq}
    \begin{split}
        \operatorname{Pr}[X \leq \lambda] \leq \exp \left(-2 n\left(\frac{K}{N}-\frac{\lambda}{n}\right)^{2}\right).
    \end{split}
    \end{equation}
    Let also $\lambda > 0$ be a positive value.
    Using Serfling's bound for the hypergeometric distribution, it holds that
    \begin{equation}
    \begin{split}
        \operatorname{Pr}\left[\sqrt{n}\left(\frac{X}{n}-\frac{N}{K}\right) \geq \lambda\right] \leq \exp \left(-\frac{2 \lambda^{2}}{1-\frac{n-1}{N}}\right).
    \end{split}
    \end{equation}
    As a corollary, we obtain the concentration inequality of hypergeometric distribution symmetric to Eq.~\eqref{eq:ineq_tail_hypergeometric_leq},
    \begin{equation}  \label{eq:ineq_tail_hypergeometric_geq}
    \begin{split}
        \operatorname{Pr}[X \geq \lambda] \leq \exp \left(-2 n\left(\frac{K}{N}-\frac{\lambda}{n}\right)^{2}\right).
    \end{split}
    \end{equation}
\end{lemma}

\section{Composable security of delegated quantum computation protocols\label{sec:appendix_preliminaries}}

Delegated quantum computation protocols~\cite{broadbent2009universal, fitzsimons2017unconditionally, leichtle2021verifying} allow clients with limited quantum capabilities, such as single-qubit state preparation and communication, to delegate tasks to a powerful server while retaining security guarantees such as blindness and verifiability. The security of these protocols can be rigorously analysed within the Abstract Cryptography (AC) framework~\cite{Maurer2011abstract}, which formalises composable security in a modular way. This modular framework enables composable security guarantees without requiring incremental and exhaustive proofs of the entire protocol whenever individual components are combined.

In what follows, we first introduce the AC framework, then review the universal blind quantum computation (UBQC) protocol~\cite{broadbent2009universal} based on MBQC, and explain how AC captures its security. Finally, we review the Verifiable Blind Quantum Computation (VBQC) protocol~\cite{fitzsimons2017unconditionally} and the Robust VBQC (RVBQC) protocol~\cite{leichtle2021verifying} combined with their security guarantees.

\subsection{Abstract cryptography (AC)\label{sec:Abstract_cryptography}}

Abstract cryptography is a cryptographic framework designed to be top-down and axiomatic to analyse the security of a protocol in an arbitrarily adversarial environment. In contrast to the conventional game-based security that analyses each specific adversarial scenario, the AC framework provides universal composable security. Composably secure protocols within the AC framework will maintain their security when composed with each other in parallel or in series, ensuring a modular composition of security for the entire combined protocol as well.

The AC framework consists of abstract systems with well-distinguished and labelled interfaces to transmit information to other systems. Systems are classified into resources, converters, filters, and distinguishes. The AC framework aims to construct a new secure resource $\pi\mathcal{R}$ from an available resource $\mathcal{R}$ and a protocol $\pi$ by showing the security of $\pi$. Here, the resource $\mathcal{R}$ is an abstract system with an index set of interfaces $\mathcal{I}$ for mediating transcripts. The protocol $\pi=\{\pi_{i}\}_{i\in\mathcal{I}}$ is a set of converters $\pi_{i}$ indexed by $\mathcal{I}$, where each converter is a two-interface system mediating between the resource and an external party.

A protocol $\pi$ is proved to be secure by showing the statistical indistinguishability between the constructed resource $\pi\mathcal{R}$ and the ideal resource $\mathcal{S}$, i.e. any distinguisher cannot distinguish with high probability the two resources $\pi\mathcal{R}$ and $\mathcal{S}$. In concrete terms, the distinguisher is an abstract system that interacts with a resource, attempting to decide whether it is connected to a real resource or an ideal one. It may send inputs, receive outputs, and exploit any observable behaviour in order to distinguish the two resources. Ultimately, the distinguisher must output a single bit indicating its guess: for instance, outputting $1$ if the distinguisher believes it is interacting with the constructed resource $\pi\mathcal{R}$ and $0$ otherwise. The formal definition of statistical indistinguishability between two resources can be stated as follows.

\begin{definition}[Statistical Indistinguishability of Resources] Let $\epsilon>0$, and let $\mathcal{R}_{1}$ and $\mathcal{R}_{2}$ be two resources with the same input and output interfaces. The resources are $\epsilon$-statistically-indistinguishable if, for any unbounded distinguisher $\mathcal{D}$, the following holds:
  \begin{equation}\label{eq:distinguishing_advantage}
    \begin{split} \left|\operatorname{Pr}\left[\mathcal{D}\left(\mathcal{R}_{1}\right)=1\right] - \operatorname{Pr}\left[\mathcal{D}\left(\mathcal{R}_{2}\right)=1\right]\right| \leq \epsilon,
    \end{split}
  \end{equation} which is denoted by $\mathcal{R}_{1}\approx_{\epsilon}\mathcal{R}_{2}$, and $\epsilon$ is referred to as distinguishing advantage.
\end{definition}

Here, the distinguishing advantage $\epsilon$ quantifies how much better a distinguisher can perform than random guessing. If two resources are completely indistinguishable, the success probability is $\displaystyle \frac{1}{2}$ (the same as random guessing), yielding $\epsilon=0$. Otherwise, the distinguishing advantage is $\epsilon$, the distinguisher can succeed with probability $\displaystyle\frac{1}{2}+\epsilon$.

\begin{figure}[htbp] \centering \subfloat[correctness: $\pi\mathcal{R}\approx_{\epsilon}\mathcal{S}$\label{fig:ac_correctness}]{ \includegraphics[width=\linewidth]{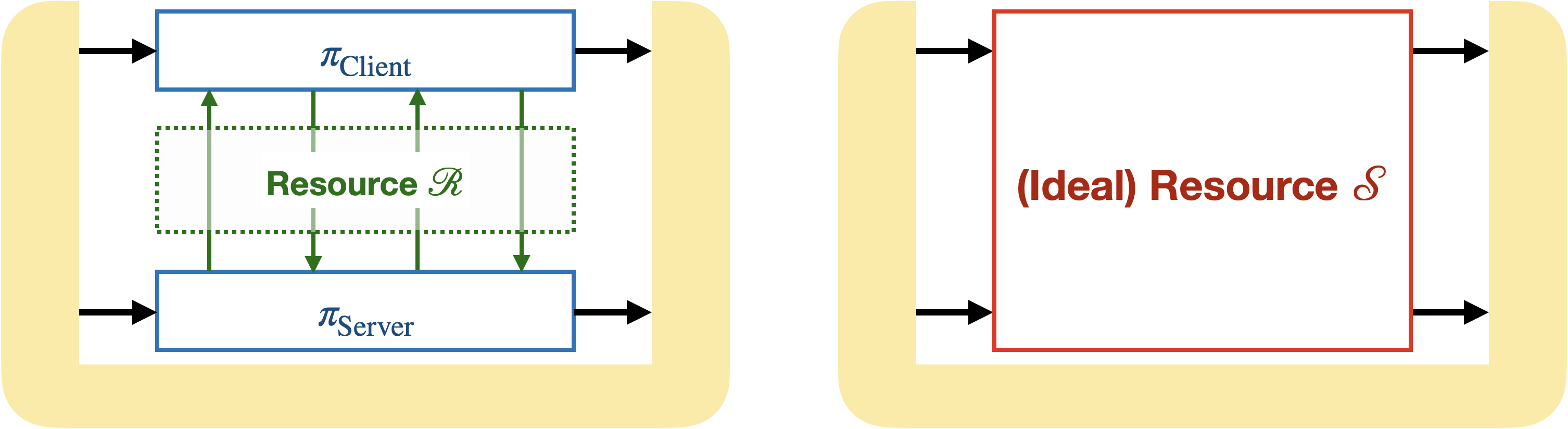} } \hfill \subfloat[security: $\pi_{\mathrm{Client}}\mathcal{R}\approx_{\epsilon}\mathcal{S}\sigma$\label{fig:ac_security}]{ \includegraphics[width=\linewidth]{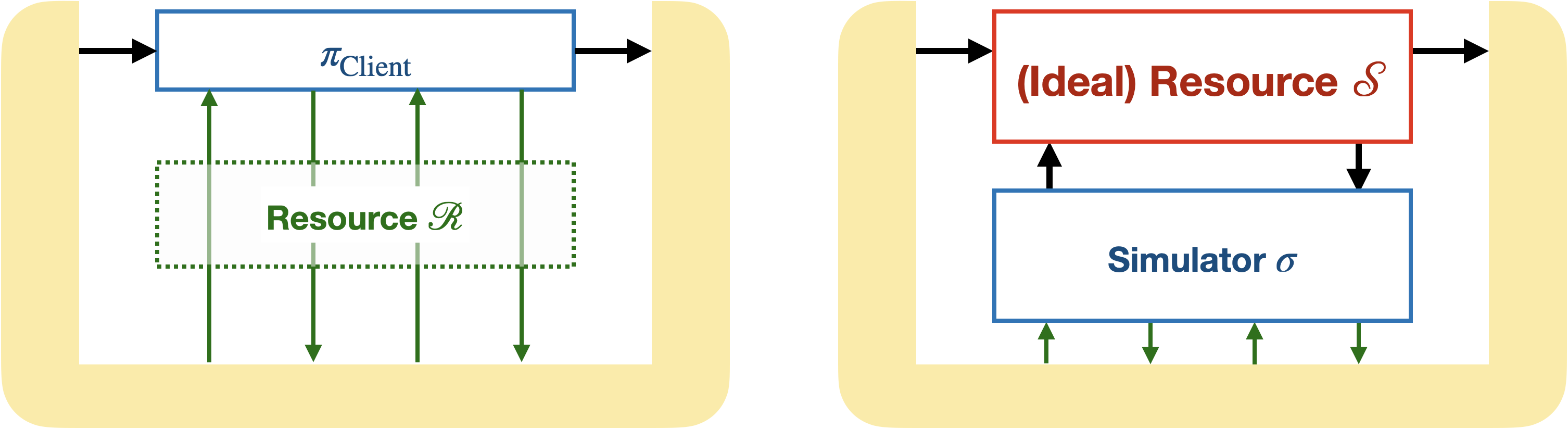} }
    \caption{The schematic illustrations of correctness and security are depicted in (a) and (b), respectively. On the basis of a secure resource $\mathcal{R}$ as an established channel between the Client and the Server, the protocol $\pi=(\pi_{\mathrm{Client}}, \pi_{\mathrm{Server}})$ constructs a new resource $\pi\mathcal{R}=\pi_{\mathrm{Client}}\mathcal{R}\pi_{\mathrm{Server}}$. The transcripts are written as arrows, and the yellow object is the distinguisher that manages the input and output transcripts between the resource and the protocol of interest. }
    \label{fig:abstract_cryptography}
\end{figure}

When constructing a resource $\pi\mathcal{R}$ from a resource $\mathcal{R}$ and a protocol $\pi$, the security of $\pi$ is then characterised by the indistinguishability between $\pi\mathcal{R}$ and the ideal resource $\mathcal{S}$. Here, we restrict to the two-party case with an honest ``Client'' and a potentially malicious ``Server''. The following definition defines how well the protocol $\pi$ constructs $\mathcal{S}$ from $\mathcal{R}$.
\begin{definition}[Construction of Resources]
  \label{def:Construction_of_Resources} Let $\epsilon>0$. We say that a two-party protocol $\pi$, between an honest Client and a potentially malicious Server, $\epsilon$-statistically-constructs resource $\mathcal{S}$ from resource $\mathcal{R}$ if,
  \begin{itemize}
      \item it is correct: $\pi\mathcal{R}\approx_{\epsilon}\mathcal{S}$, i.e. when the Server is honest, the client-side outputs between $\pi\mathcal{R}$ and $\mathcal{S}$ are $\epsilon$-statistically indistinguishable;
      \item it is secure against the malicious Server, i.e. there exists a simulator $\sigma$ such that $\pi_{\mathrm{Client}}\mathcal{R}\approx_{\epsilon}\mathcal{S}\sigma$, where $\pi_{\mathrm{Client}}$ is $\pi$'s Client side protocol.
  \end{itemize}
\end{definition}

Intuitively, correctness ensures that the protocol behaves as intended when all parties are honest, while security guarantees that malicious behaviour can be emulated in the ideal world by a simulator, thereby preserving composable security. The existence of such a simulator implies that the use of $\pi\mathcal{R}$ with a malicious Server is still well-indistinguishable from using the ideal resource $\mathcal{S}$, which is designed to be secure. The schematic illustrations of correctness and security in Definition~\ref{def:Construction_of_Resources} are presented in Fig.~\ref{fig:abstract_cryptography}.

Using the definitions above, we can state the following general composition theorem~\cite{Maurer2011abstract} that guarantees the additive accumulation of distinguishing advantage when composing two statistically secure protocols.
\begin{theorem}[General Composition of Resources~\cite{Maurer2011abstract}] Let $\mathcal{R}$, $\mathcal{S}$ and $\mathcal{T}$ be resources, $\alpha, \beta$ and $\mathsf{id}$ be protocols, where protocol $\mathsf{id}$ does not modify the resource it is applied to. Let $\circ$ and $|$ denote the sequential and parallel composition of protocols and resources, respectively. Then the following implications hold:
    \begin{itemize}[left=5pt]
        \item Sequential composability: \\ if $\alpha \mathcal{R} \approx_{\epsilon_{\alpha}} \mathcal{S}$ and $\beta\mathcal{S} \approx_{\epsilon_{\beta}} \mathcal{T}$, then $\left(\beta \circ\alpha\right) \mathcal{R} \approx_{\epsilon_{\alpha}+\epsilon_{\beta}}\mathcal{T}$.
        \item Context insensitivity: \\ if $\alpha \mathcal{R} \approx_{\epsilon_{\alpha}} \mathcal{S}$, then $\left(\alpha \mid \mathrm{id}\right)\left(\mathcal{R} \mid \mathcal{T}\right) \approx_{\epsilon_{\alpha}} \left(\mathcal{S} \mid \mathcal{T}\right)$.
    \end{itemize} Combining these two properties yields the composability of protocols.
\end{theorem}

\subsection{Universal blind quantum computation (UBQC)}

\begin{resource}[htbp]
    \caption{\raggedright Blind Delegated Quantum Computation (BDQC)}
    \label{res:bdqc}
    \begin{algorithmic}[0] \STATE \textbf{Public Information:} $\left(\mathfrak{C}, G, f\right)$ defined as below. \STATE \textbf{Inputs at the Client's interface:} The target computation $\mathsf{C}\in \mathfrak{C}$, its associated measurement pattern $C$ that contains the graph $G=(V,E)$, the input and output sets $\mathrm{I}, \mathrm{O}\subseteq V$, the measurement angles $\{\phi_{v}\}_{v\in V}$, and the flow $f$. \STATE \textbf{Process at the Server's interface:}
        \begin{enumerate}
            \item The Resource receives from the Server $e\in\{0,1\}$, a flag whether to leak information to Server.
            \item If $e=1$, the Resource sends to the Server the allowed leakage $l_{\mathfrak{C}} = \left(\mathfrak{C}, G, f\right)$.
            \item The Resource receives at its Server's interface the deviation $\left(\rho_{\mathrm{R}}, \mathsf{F}\right)$ as a pair of an ancillary state $\rho_{\mathrm{R}}$ and a CPTP map $\mathsf{F}$.
        \end{enumerate} \STATE \textbf{Outputs at the Client's interface:} The Resource sets $\vec{\mathrm{b}} := \operatorname{Tr}\left[\mathsf{F}\left(|\vec{b}\rangle\langle\vec{b}|\otimes\rho_{\mathrm{R}}\right)\right] \in\{0,1\}^{|\mathrm{O}|}$, where $\vec{b}\in\{0,1\}^{|\mathrm{O}|}$ is the correct output following the procedure of measurement pattern $C$ corresponding to the target computation $\mathsf{C}$. The Resource returns $\vec{\mathrm{b}}$ at the Client's interface.
    \end{algorithmic}
\end{resource}

The UBQC protocol achieves perfect blindness for the Client to delegate its quantum computation to the untrusted Server, when the Client can prepare and send a sequence of single qubits through quantum communication. The procedure of UBQC is based on the measurement-based quantum computation (MBQC) model grounded in the principle of gate teleportation~\cite{gottesman1999demonstrating, raussendorf2001a, knill2001a, danos2007the, briegel2009measurement-based}. In this model, the computation proceeds by first preparing a highly entangled resource state, typically a graph state, and then performing a sequence of adaptive single-qubit measurements in rotated bases. The measurement outcomes determine subsequent measurement angles, enabling the realisation of arbitrary quantum operations. More formally, the procedure of MBQC is defined as the following measurement pattern.

\begin{definition}[Measurement Pattern] A pattern in the MBQC model is given by a graph $G = (V, E)$, input and output vertex sets $\mathrm{I}, \mathrm{O}\subseteq V$, a flow function $f$ which induces a partial order $\preceq_{G}$ of the qubits $V$, and a set of measurement angles $\left\{\phi_{v}\right\}_{v\in V}$ in the $X$-$Y$ plane of the Bloch sphere.
\end{definition}

\begin{protocol}[H]
    \caption{\raggedright Universal Blind Quantum Computation (UBQC)}
    \label{proto:ubqc}
    \begin{algorithmic}[0] \STATE \textbf{Inputs from Client:} The target computation $\mathsf{C}\in \mathfrak{C}$, its associated measurement pattern $C$ that contains the graph $G=(V,E)$, the input and output sets $\mathrm{I}, \mathrm{O}\subseteq V$, the measurement angles $\{\phi_{v}\}_{v\in V}$, and the flow $f$. \STATE \textbf{Protocol:}
        \begin{enumerate}
            \item The Client sends the graph's description $\left(G, \mathrm{I}, \mathrm{O}\right)$ to the Server.
            
            \item The Client generates secret parameters:
            \begin{enumerate}
                \item ($\mathsf{X}$ randomisation) The Client chooses a random bit $\mathrm{a}_{v}^{\mathrm{init}}\in\{0,1\}$ for $v\in \mathrm{I}$ and sets $\mathrm{a}_{v}^{\mathrm{init}}=0$ for $v\in V\setminus \mathrm{I}$. The Client also computes $\displaystyle \mathrm{a}_{v}^{\mathrm{prop}} = \bigoplus_{j\in N_{G}(v)}\mathrm{a}_{j}^{\mathrm{init}} \in\{0,1\}$ for all $v\in V$.
                \item ($\mathsf{Z}$ randomisation) The Client chooses a random bit $\mathrm{r}_{v}\in\{0,1\}$ for all $v\in V$.
                \item (randomisation for blindness) The Client chooses a random $\theta_{v}\in\Theta$ for all $v \in V$.
            \end{enumerate}
            
            \item The Client prepares and sends to the Server all single qubits for $v\in V$. For $v\in \mathrm{I}$, the Client sequentially sends each qubit in $\displaystyle \left(\bigotimes_{v\in\mathrm{I}}\mathsf{Rz}_{v}(\theta_{v})\mathsf{X}_{v}^{\mathrm{a}_{v}^{\mathrm{init}}}\right)\left[\rho_{\mathrm{init}}\right]$. For $v\in V\setminus\mathrm{I}$, the Client sends $\displaystyle |+_{\theta_{v}}\rangle$.
            
            \item The Server applies a $\mathsf{CZ}$ gate between qubits $v_{1}$ and $v_{2}$ if $(v_{1}, v_{2})\in E$.
            
            \item For each $v\in V$, the Client and Server interactively perform the MBQC process. Once the Client receives the measurement outcome $\mathrm{b}_{j}\in\{0,1\}$ for all $j\in S_{X,v}\cup S_{Z,v}$, where $S_{X,v} = f^{-1}\left(v\right), S_{Z,v} = \left\{j~|~v\in N_{G}\left(f\left(j\right)\right)\right\}$, the Client computes the adaptive angle update $\phi_{v}^{\prime}$. The Client then computes the measurement angle $\delta_{v}$ masked with $\mathrm{a}_{v}^{\mathrm{init}}$, $\mathrm{a}_{v}^{\mathrm{prop}}$, and $\mathrm{r}_{v}$ for the QOTP randomisation, and $\theta_{v}$ for the blindness:
            \begin{equation}\label{eq:delta_UBQC}
            \begin{split} \quad\quad\quad\quad s_{X,v} &= \bigoplus_{j \in S_{X,v}} \mathrm{b}_{j} \oplus \mathrm{r}_{j}, \quad s_{Z,v} = \bigoplus_{j \in S_{Z,v}} \mathrm{b}_{j} \oplus \mathrm{r}_{j}, \\ \phi_{v}^{\prime} &= (-1)^{s_{X,v}} \phi_{v}+s_{Z,v} \pi, \\ \delta_{v} &= \left(-1\right)^{\mathrm{a}_{v}^{\mathrm{init}}} \phi_{v}^{\prime} + \theta_{v}+\left(\mathrm{r}_{v} + \mathrm{a}_{v}^{\mathrm{prop}}\right) \pi.
            \end{split}
            \end{equation} Note that $s_{X,v} = s_{Z,v} = 0$ for $v\in \mathrm{I}$. The Client sends to the Server the angle $\delta_{v}$ and the Server returns to the Client a bit $\mathrm{b}_{v}\in\{0,1\}$ as a measurement result of qubit $v$ with basis $\{|+_{\delta_{v}}\rangle, |-_{\delta_{v}}\rangle\}$.
            
            \item The Client returns $\vec{\mathrm{b}}\oplus\vec{\mathrm{r}}\in\{0,1\}^{|\mathrm{O}|}$ as the final output, where $\vec{\mathrm{b}}\in\{0,1\}^{|\mathrm{O}|}$ (resp. $\vec{\mathrm{r}}$) is a bitstring of the binaries $\mathrm{b}_{v}$ (resp. $\mathrm{r}_{v}$) for all $v\in\mathrm{O}$.
            
        \end{enumerate}
    \end{algorithmic}
\end{protocol}

To rigorously define the task of blind delegation of quantum computation, we introduce the Blind Delegated Quantum Computation (BDQC) resource, where, by design, the server never learns the precise computation but instead only the class of computation that the delegated task belongs to $\mathfrak{C}$, i.e. the prepared graph $G$ and its flow $f$. This resource can then be constructed perfectly using the UBQC protocol~\cite{broadbent2009universal}, described in Protocol~\ref{proto:ubqc}. Note that Protocol~\ref{proto:ubqc} is adapted to classical outputs only.

One can describe the composable security of the UBQC protocol by the words of the AC framework. To state this, one first defines an ideal resource as a hypothetical system that achieves the desired functionality, which is secure by definition. For the case of UBQC, the ideal functionality returns potentially biased output by keeping the blindness of the computation up to the allowed leakage $l_{\mathfrak{C}} = (\mathfrak{C}, G, f)$. This is formally defined as the BDQC resource (Resource~\ref{res:bdqc}) that enables the server to influence the outcome by modelling a potential deviation, while leaking no information to the server beyond the prescribed nature of leakage. The security is then analysed by evaluating the indistinguishability between the UBQC protocol and the BDQC resource with respect to the correctness and security against the malicious Server defined in Definition~\ref{def:Construction_of_Resources}.

The UBQC protocol is shown to achieve perfect composable security~\cite{dunjko2014composable}, i.e. the UBQC protocol and the BDQC resource are perfectly indistinguishable. The blindness, which is the only functionality of interest in the BDQC resource, is realised by the Client's use of the quantum one-time pad (QOTP)~\cite{knill2004fault-tolerant, kern2005quantum, dankert2009exact, geller2013efficient, wallman2016noise} to randomise measurement angles and measurement results. Formally, the security of UBQC is stated as follows.

\begin{theorem}[Security of UBQC~\cite{dunjko2014composable}]\label{theorem:UBQC} The UBQC protocol (Protocol~\ref{proto:ubqc}) perfectly constructs the BDQC resource (Resource~\ref{res:bdqc}) leaking only public information $\left(\mathfrak{C}, G, f\right)$.
\end{theorem}

\begin{resource}[htbp]
    \caption{\raggedright Secure Delegated Quantum Computation (SDQC)}
    \label{res:sdqc}
    \begin{algorithmic}[0] \STATE \textbf{Public Information:} $\left(\mathfrak{C}, G, f, N\right)$ defined as below. \STATE \textbf{Inputs at the Client's interface:} The target computation $\mathsf{C}\in \mathfrak{C}$, its associated measurement pattern $C$ that contains the graph $G=(V,E)$, the input and output sets $\mathrm{I}, \mathrm{O}\subseteq V$, the measurement angles $\{\phi_{v}\}_{v\in V}$, the flow $f$, and the number of total rounds $N$. \STATE \textbf{Process at the Server's interface:}
        \begin{enumerate}
            \item Receive from Server $e\in\{0,1\}$, a flag whether to leak information to Server.
            \item If $e=1$, send to Server the allowed leakage $l_{\mathfrak{C}}=\left(\mathfrak{C}, G, f, N\right)$.
            \item Receive from Server $d\in\{0,1\}$, a flag whether to deviate the computation.
        \end{enumerate} \STATE \textbf{Outputs at the Client's interface:} Let $\vec{b}\in\{0,1\}^{|\mathrm{O}|}$ be the correct output following the procedure of measurement pattern $C$ corresponding to $\mathsf{C}$.
        \begin{enumerate}
            \item If $d=0$, set $\vec{\mathrm{b}} := \vec{b}$ and return $(\mathsf{Acc}, \vec{\mathrm{b}})$ to the Client.
            \item If $d=1$, return $(\mathsf{Rej}, \perp)$.
        \end{enumerate}
    \end{algorithmic}
\end{resource}

\begin{protocol}[htbp]
    \caption{\raggedright Robust VBQC (RVBQC)~\cite{leichtle2021verifying}}
    \label{proto:rvbqc}
    \begin{algorithmic}[0] 
    \STATE \textbf{Inputs from Client:} The target computation $\mathsf{C}\in \mathfrak{C}$, the graph $G=(V,E)$, the flow $f$, and the $K$-colouring $\left\{\mathsf{V}_{k}\right\}_{k=1}^{K}$ of $G$. 
    \STATE \textbf{Protocol:}
        \begin{enumerate}
            \item The Client randomly samples indices in $[N]$ for $N = N_{c} + N_{t}$ to indicate the locations of test and computation rounds. Let $\mathrm{S}_{\mathsf{T}}$ (resp. $\mathrm{S}_{\mathsf{C}}$) be the index set of test (resp. computation) rounds.
            
            \item For $i \in \mathrm{S}_{\mathsf{T}}$, the Client constructs a test round following the same procedure as for the test rounds of the RVBQC (Protocol~\ref{proto:rvbqc}).
            \begin{enumerate}
                \item The Client chooses uniformly at random a colour $\mathsf{V}_{j}\in\left\{\mathsf{V}_{k}\right\}_{k=1}^{K}$ to specify the trap qubits.
            
                \item The Client sends qubits to the Server. If $v \notin \mathsf{V}_{j}$ (dummy), the Client chooses a bit $\mathrm{d}_{v} \in \{0,1\}$ uniformly at random and sends the state $|d_{v}\rangle$. Otherwise, the Client chooses $\theta_{v}\in\Theta$ at random and sends the state $|+_{\theta_{v}}\rangle$.
                
                \item The Server performs $\mathsf{CZ}$ gates between all its qubits corresponding to an edge in the set $E$.
                
                \item For $v \in V$, the Client sends a measurement angle $\delta_{v}$, the Server measures the appropriate corresponding qubit in the basis $\{|+_{\delta_{v}}\rangle, |-_{\delta_{v}}\rangle\}$, returning outcome $\mathrm{b}_{v}$ to the Client. The angle $\delta_{v}$ is defined as follows:
                \begin{itemize}
                    \item If $v \notin \mathsf{V}_{j}$ (dummy): the Client chooses $\delta_{v}$ from $\Theta$ uniformly at random.
                    \item If $v \in \mathsf{V}_{j}$ (trap): the Client chooses $\mathrm{r}_{v} \in\{0,1\}$ uniformly at random and sets $\delta_{v} = \theta_{v} + \mathrm{r}_{v} \pi$.
                \end{itemize}
                    
                \item For all $v \in \mathsf{V}_{j}$ (traps), the Client computes $\displaystyle \mathrm{d}_{v}=\bigoplus_{k \in N_{G}(v)} d_{k}\in\{0,1\}$, the sum over the values of neighbouring dummies of the trap qubit $v$ in the $i$th round. the Client then verifies whether $\mathrm{b}_{v}=\mathrm{r}_{v} \oplus \mathrm{d}_{v}$ holds for all $v \in \mathsf{V}_{j}$. If this does not hold, the test round is considered failed.
            \end{enumerate}

            \item For $i \in \mathrm{S}_{\mathsf{C}}$, the Client delegates to the Server the computation round using UBQC (Protocol~\ref{proto:ubqc}).

            \item If more than $\omega N_{t}$ test rounds failed, the Client returns $\mathsf{Abort}$. 
            Otherwise, the Client performs a majority vote over the outputs of the computation rounds: if there exists an outcome that appears in more than half of the computation rounds, the Client returns that outcome; otherwise, the Client returns $\mathsf{Abort}$.
            
        \end{enumerate}
    \end{algorithmic}
\end{protocol}

\subsection{Robust verifiable blind quantum computation (RVBQC)}

While the UBQC protocol ensures blindness, the client may also wish to verify the result of the computation, i.e. ensure that the provided result has not been tempered with. 
This is expressed by the Secure Delegated Quantum Computation (SDQC) resource in Resource~\ref{res:sdqc}.

Clearly, the SDQC resource either provides the expected result or aborts depending on the flag bit $d$ transmitted by the malicious server. The Verifiable Blind Quantum Computation (VBQC) protocol~\cite{fitzsimons2017unconditionally} constructs the SDQC resource with negligible distinguishability by embedding ``trap'' qubits and ``dummy qubits'' into the UBQC protocol so that it can both execute the computation while probing the behaviour of the server.

More precisely, the trap qubits are single-qubit deterministic computations whose outcomes are efficiently simulable by the Client while remaining hidden from the Server. Specifically, a trap qubit is initialised into $|+_{\theta}\rangle$ and measured on the $\{|+_{\theta}\rangle, |-_{\theta}\rangle\}$ basis, outputting the eigenvalue $1$ under its honest execution. The dummy qubits serve to isolate the trap qubits on the MBQC pattern and to mask the locations of the traps. Note that our framework is not restricted to isolated trap qubits according to the graph colouring, and one can also use further optimised dummyless traps~\cite{Kapourniotis2025Asymmetric} and general traps~\cite{kapourniotis2024unifying}.

With this trappification scheme, the Client can detect deviations if a trap is affected by a harmful deviation that can non-trivially affect the computation. 
However, the embedded traps and dummies among computation qubits would increase the qubit overhead as they enlarge the MBQC pattern. 
Besides, VBQC requires fault tolerance for the security amplification to achieve a negligible construction error in the AC framework.

These issues are solved for BQP computations by the Robust VBQC (RVBQC) protocol~\cite{leichtle2021verifying}, keeping the verifiability guarantees of VBQC while leveraging a minimal overhead construction and introducing robustness against noise.
As described in Protocol~\ref{proto:rvbqc}, the protocols consist of $N_{c}$ computation rounds and $N_{t}$ test rounds containing trap and dummy qubits.
A threshold $0<\omega<1$ is introduced to allow a constant number of test failures without aborting the computation, thus ensuring robustness against noise. 
For the computation rounds, the protocol repeats the delegated task multiple times and applies a classical majority vote over the outputs, enabling the Client to classically amplify correctness and achieve negligible construction error for the SDQC resource in the AC framework~\cite{leichtle2021verifying, kapourniotis2024unifying}.

\bibliography{main}

\end{document}